# The Fair Basis

**Funding and capital in the reduced form framework**

Wujiang Lou[1]

August 28, 2017; Updated April 24, 2019

**Abstract**

A negative basis trade enters a long bond position and buys protection on the issuer of the bond through credit default swap (CDS), aiming at arbitrage profit due to the bond-CDS basis. To classic reduced form model theorists, the existence of the basis is an abnormality or merely liquidity noise. Such a view, however, fails to explain large basis trading losses incurred during the financial crisis. Employing a bond continuously hedged by CDS under a dynamic spread model with bond repo financing, we find that there is unhedged and unhedgeable residual jump to default risk that can't be diversified because of credit correlation. An economic capital approach has to apply and a charge on the use of capital follows. Together with the hedge funding cost, it allows us to better understand the basis's economics and to predict its fair level.

**Keywords:** CDS-bond basis, negative basis, reduced form model, default risk, hedging error, FVA, KVA.

## 1. Introduction

A negative basis trade enters a long bond position and buys protection on the issuer of the bond through credit default swap (CDS), aiming at profit due on a positive carry when the bond's funding cost is lower than the bond-CDS basis. Prior to the 2007-2009 financial crisis, basis trading was a popular credit arbitrage trading strategy that excited many hedge funds and banks' prop trading desks. At least to some, it had turned out to be elusive, when significant losses[2] were unfolded. The basis, defined as the CDS spread minus same maturity bond spread, reflects a pricing discrepancy between the bond market and the CDS market on the same credit risk. Prior to the financial crisis, it was not a large number, rarely exceeding 20 bps. The financial crisis, however, saw an unprecedented basis widening, see Figure 1, with investment grade (IG) basis spiked at 250 bps and high yield (HY) at 650 bps.

---

[1] The views and opinions expressed herein are strictly the views and opinions of the author, and do not reflect those of his employer and any of its subsidiaries. The author wishes to thank Joseph Langsam and Simon Juen for helpful comments.

While there are certain factors such as CDS's cheapest-to-delivery option, lack of voting rights conveyed to bond holders, protection seller's credit risk, and distortion created by LIBOR discounting (D.E. Shaw 2009), that could contribute to the basis, earlier empirical researches (e.g., Longstaff, Mithal, and Neis, 2005) attribute the basis mostly to relative liquidity between bond and CDS markets. Such is the case, the basis is considered non-economic and often termed the liquidity basis. At the time, CDS was generally considered of better market liquidity. Following the crisis, CDS trading has been shrinking, due to tightened regulatory scrutiny and capital rules. Many banks exited from single name CDS trading (Burne and Henning, 2014). At the same time, the push for electronic trading and trade data collection has improved cash bond liquidity.

In the industry, however, it has been prominently characterized as a funding basis, and the negative basis trade is often considered a funding arbitrage trade as credit risk is perfectly hedged. Bai and Collin-Dufresne (2013)'s empirical study finds that funding liquidity, the ability to roll short term funding, plays a significant role in explaining historical basis, confirming the industry's understanding and experience.

On the modeling side, the reduced form model with issuer default probability calibrated to the CDS curve would predict a bond fair price quite different from the bond market price. The reduced form credit model (Jarrow and Turnbull, 1992 & 1995; Lando, 1998; Duffie and Singleton, 1999) extends the interest rate term structure modeling to default modeling but does not consider funding and other factors when pricing a risky bond. In the risk neutral pricing theory, once a default time model is put in place, aggregation and discounting of default contingent cash flows lead to a net present value (npv). A default intensity process is assumed and default happens as the first jump of a Cox process. The existence of the risk neutral measure is provided for by assuming an arbitrage-free CDS market, which is used to calibrate the intensity process. The other approach is, of course, the structural or options approach (Merton 1974), which treats debt and equity as two complimentary contingent claims on the same firm asset value. Because the firm asset is not tradeable and its value can only be inferred from the equity value, this options based approach[3], however, does not follow the standard Merton (1976) dynamic hedging scheme.

---

[2] According to Fontana (2010), Deutsche Bank's prop credit trading unit reportedly lost one billion in 2007 when some credits' basis widened in the wrong direction, Merrill Lynch lost multiple billions in 2008 around the time when it was bought by bank of America, and in early 2008, Citadel's flagship investment fund was down significantly due to failed basis trades.

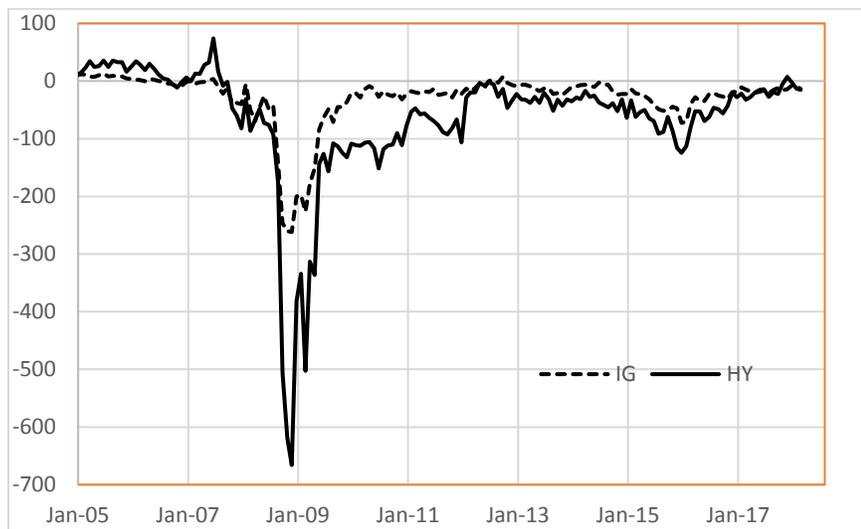

Figure 1. Historical IG and HY corporate CDS-bond monthly basis (bps, indices from JP Morgan Research) from Jan 2005 to March 2018.

Recall that the classic Black-Scholes-Merton economy is complete. An option is attained by dynamic trading in its underlying stock. In other word, the option can be hedged without any error. In the credit market, such an exercise can't be done generally without incurring hedging errors. Credit derivatives pricing models are therefore considered incomplete market models. Nonetheless, pricing can procced as no-arbitrage pricing is perceived to be guaranteed by calibrating to CDS markets. In some sense, credit models can be understood to have implicitly priced in credit market incompleteness by resorting to the invisible hand of the CDS market.

---

[3]Debt and equity are hedgeable at least theoretically when the firm value is modelled as a diffusion process. The default time however becomes predictable. To relax that, random debt barrier is introduced or jumps are added to the firm value. But once we do that, the economy is incomplete again. This line of models either over-predict credit spreads for high yield bonds or under-predict spreads for investment grade bonds (Eom 2005). It remains popular as a default probability model for credit risk management purposes.

Note that in Merton (1976), stocks' jumps are considered idiosyncratic and not hedgeable. Merton resorts to the hedging portfolio's diversification argument under the Arbitrage Pricing Theory (APT) to derive a mixed differential-difference Black-Scholes type equation and an option pricing formula with lognormally distributed jumps. In the credit market, credits are known of correlated defaults (Duffie et al 2007). A default model then separates into systemic (common) factors and idiosyncratic factors. This presents two issues: first, the common factor is not hedgeable, and secondly the idiosyncratic factors are of finite diversification, not infinite. The second issue can always be put aside as a merely theoretical concern, for in practice, every portfolio is finite but can always be thought of a good approximation of an infinite market. With regard to the first issue, typically a limited number of agents will warehouse the systemic risk. In doing so, a necessary compensation involves a charge on the economic capital incurred.

Lou (2016) explores an economic capital approach to derivatives hedging error: the delta hedging strategy is designed to zero out mean hedging error while a capital reserve is taken as the error's VaR measure. The fair value of the derivatives then allows a compensation for the cost of putting up the capital reserve. The Black-Scholes-Merton option pricing framework is extended and applied to the gap risk embedded in repurchase agreements (repo). At zero haircut, a one-year repo with 10 day MPR on main equity could command capital charges as large as 50 bp per annum for a 'BBB' rated borrower. Increased haircut reduces capital charges, e.g., to 4 bp at 10% haircut. This shows that hedging error or gap risk contributes to the basis between the repo rate and the risk-free rate. Naturally we would expect the same factor contributing to the bond-CDS basis.

This article contributes to the literature by deriving a model of credit risky or defaultable bond valuation under economic capital cost and funding as well. The approach taken deviates from the risk neutral pricing model: we instead turn to a bond-CDS hedging economy, to incorporate funding cost and identify hedging error. This paper's innovation is to associate the hedging error with economic capital and build a cost of capital into the valuation model. The fair value is different from the risk-neutral price of the bond, resulting in a bond value adjustment (xva). This xva becomes the *fair* basis, when converted to a running spread per annum, which can be used as a relative value measure for the basis trading strategy.

**2. Review of Credit Default Swaps Pricing**

Let $\lambda(t)$ be the default intensity of a Cox process defined in a probability space $(\Omega,P,F)$. A bond issuer C defaults when the Cox process has its first jump at time $\tau$. Let $\Gamma_t = I[\tau \leq t]$ denote the default indicator, 1 if default before time $t$, 0 otherwise. Then the survival probability at time $T$ as seen from time $t$ denoted by $q_t(T)$ is

$$q_t(T) = \Pr(\Gamma_T = 0) = E_t^Q[e^{-\int_t^T \lambda(u)du}] \qquad (1)$$

where $E^Q$ is an expectation taken under the risk-neutral measure $Q$. Let $\beta(t)$ be the applicable discount factor, $\beta(t) = e^{-\int_0^t r(u)du}$ with $r(t)$ being the riskfree rate, the net present value (*npv*) of the loss process $l(t)$ due to default, or default present value (*dpv*), is given by

$$dpv = E^Q\left[\int_0^T \beta(t)dl(t)\right] = E^Q\left[\int_0^T \beta(t)(1-R)e^{-\int_0^t \lambda(u)du}\lambda(t)dt\right] \qquad (2)$$

where $R$ is the recovery rate on notional. The present value of a unit spread on a unit notional, or annuity denoted by *apv*, is

$$apv = E^Q\left[\int_0^T e^{-\int_0^t (r(u)+\lambda(u))du}dt\right] \qquad (3)$$

The net present value of the CDS is *npv = dpv – S\*apv*, where *S* is the CDS premium[4].

A fixed coupon risky bond is decomposed as an annuity leg, a principal leg subject to the issuer's survival at maturity, and a short protection or loss leg. Credit linked notes, total return swaps and other credit derivatives products can be decomposed similarly.

Assuming that the intensity is governed by a diffusion process $d\lambda_t = adt + bdW_t$, the survival probability is given by the Feynman-Kac theorem,

---

[4] For simplicity, the premium is assumed to pay continuously. apv formula will be broken down into a sum of premium periods plus premium accrual to account for default in the mid of a period, see Lando (1998).

$$\frac{\partial q}{\partial t} + a\frac{\partial q}{\partial \lambda} + \tfrac{1}{2}b^2\frac{\partial^2 q}{\partial \lambda^2} - \lambda q = 0 \tag{4}$$

In practice, the risk neutral measure is implicitly granted as a result of calibrating to the CDS market.

### 2.1. CDS pricing with a PDE approach

We set up an economy of a long CDS protection and a bank account $M_t$. Let $V_t$ be the pre-default value of the CDS, and $\pi_t$ the economy's wealth,

$$\pi_t = M_t + (1-\Gamma_t)V_t, \tag{5}$$

Noting that the post-default value of the CDS protection is *1-R*, *R* assumed constant for ease of exposition, the self-financing equation consists of earned interest on the bank account, premium paying out, and protection payment at default,

$$dM_t = rM_t dt + (1-\Gamma_t)(-Sdt) + (1-R)d\Gamma_t \tag{6}$$

Differentiating (5) and plugging (6) to obtain

$$d\pi_t - r\pi_t dt = (1-\Gamma_t)(dV_t - rV_t dt - Sdt) + (1-R-V_t)d\Gamma_t \tag{7}$$

In the Markovian world, assuming $V(t,\lambda)$ is twice differentiable and applying Ito's lemma lead to

$$d\pi_t - r\pi_t dt = (1-\Gamma_t)dt\left(\frac{\partial V}{\partial t} + a\frac{\partial V}{\partial \lambda} + \tfrac{1}{2}b^2\frac{\partial^2 V}{\partial \lambda^2} - (r+\lambda)V - S + \lambda(1-R)\right) \\ + (1-\Gamma_t)\frac{\partial V}{\partial \lambda}bdW_t + (1-R-V)(d\Gamma_t - (1-\Gamma_t)\lambda dt) \tag{8}$$

The last two terms, a diffusion and a compensated Poisson process, are martingales. If we set the *dt* term to zero, the fair value *V* is governed by

$$\frac{\partial V}{\partial t} + a\frac{\partial V}{\partial \lambda} + \tfrac{1}{2}b^2\frac{\partial^2 V}{\partial \lambda^2} - (r+\lambda)V - S + \lambda(1-R) = 0 \tag{9}$$

And equation (8) becomes, after multiplying the deflator $\beta_t$,

$$d(\beta_t \pi_t) = (1-\Gamma_t)\beta_t b\frac{\partial V}{\partial \lambda}dW_t + (1-R-V)\beta_t(d\Gamma_t - (1-\Gamma_t)\lambda dt) \tag{10}$$

This says that the riskfree rate discounted investment portfolio $\beta_t\pi_t$ is a martingale. By applying the Feynman-Kac formulae to equation (9), noting that at expiry *T*, $V(T,\lambda_T)=0$, the risk neutral pricing formula for the CDS is obtained

$$V(t) = E_t^Q [\int_t^T e^{-\int_t^s (r(u)+\lambda(u))du} (\lambda(s)(1-R) - S) ds] \tag{11}$$

Obviously, this can be split into two parts, default present value (*dpv*) and unit premium or annuity present value (*apv*) as already given by equations (2 & 3). Similarly for a continuously paid fixed coupon bond with coupon rate $r_c$, its PDE is given by

$$\frac{\partial B}{\partial t} + a \frac{\partial B}{\partial \lambda} + \tfrac{1}{2} b^2 \frac{\partial^2 B}{\partial \lambda^2} - (r+\lambda)B + r_c + \lambda R = 0 \tag{12}$$

The survival probability PDE (4) can be obtained by noting that it is equivalent to zero payoff upon default, i.e., *R=1* in a zero interest rate environment. The risky annuity can also be solved from (9) by setting *R* to 1 and *S* to 1. Similar to equation (11), a bond pricing formulae can be obtained by applying the Feynman-Kac theorem,

$$B(t) = E_t^Q [e^{-\int_t^T (r(u)+\lambda(u))du} + \int_t^T e^{-\int_t^s (r(u)+\lambda(u))du} (\lambda(s)R + r_c) ds] \tag{13}$$

### 2.2 Hedging error

From a dynamic portfolio perspective, $\pi_t$ is a hedging error, although here no hedging is taken place. If we start the economy with $\pi_0=0$, equation (10) shows that $d(\beta_t \pi_t)$ is a martingale, so $E[\beta_t \pi_t]=0$. The deflated wealth thus behaves like a noise or a residual error with zero mean, but non-zero conditional variance. Although no risk hedging takes place in the economy, the risky cashflow is offset on an average basis: the jump to default risk is evened out by accumulating premium payments and the diffusion risk is a random walk with zero mean.

When the intensity is deterministic, the diffusion term drops out of equation (10), and the discounted wealth is a pure jump process. If we consider the bank forming an investment portfolio of credits, each having its own non-stochastic hazard rate and its default time as the first jump of a time inhomogeneous Poisson process, then these residual errors are independent. By the same diversification argument employed in Merton (1976), they will have no pricing impact. To be precise, the total error of the credit portfolio approaches zero with probability one by virtue of the Central Limit theorem. PDE (9) is justified on the basis of Ross's Arbitrage Pricing Theory (APT) under perfect diversification. The risk neutral pricing formulae (2 & 3) still stand as they are simply the necessary results of the Feynman-Kac theorem applied to PDE (9).

If the intensity is dynamic, the residual error has both a diffusion component and a pure jump component, the first and second terms of equation (10) respectively. Obviously, if these diffusions per issuer of the investment portfolio are independent, then the same diversification argument alone could lay the no-arbitrage foundation for equations (9 & 11). Credits are known to exhibit strong correlations (Duffie *et al*, 2007), however, so systemic credit risk will exist. Certain classes of the market participants are required to manage the undiversifiable residual variance in the form of provisional loss or capital. Banks and broker-dealers, for instance, are regulated and operate under stringent minimum capital requirements. As the variance is irreducible market-wide, and its management incurs costs, should the risk-neutral pricing theory be amended to price it in?

## 3. Dynamic Hedging Risky Bond with CDS

Consider a negative basis trade in which a bank's trader B buys a bond and CDS protection simultaneously to earn a riskfree carry. The trader holds a unit notional of a coupon bond issued by C priced at $B_t$, and buys protection on $\Delta_t$ unit of notional under a CDS referencing party C's credit. $B_t$ is the pre-default price, i.e., the bond price conditional on no-default prior to time $t$. Thus it is continuous. $r_c$ is the coupon rate of the bond, continuously paid. $\Delta_t$ is the bank's hedging strategy, and given our setting, $\Delta_t \geq 0$, intuitively. The trading strategy is intended to hold to maturity, so the CDS's expiry matches the bond maturity[5]. When dynamic trading is performed, we always buy or sell at the same CDS premium $S$. Additionally, we have these accounts in the economy.

- **Bank account**: The segregated economy's only investment option is a cash deposit account, with balance $M_t \geq 0$, earning the riskfree rate $r(t)$.
- **CDS margin account**: CDS is subject to full cash variation margin (VM)[6]. Let $V_t$ denote the fair value of the CDS per unit notional. $V_t$ could be negative. Further denote $L_D$ the bilateral VM cash account (unsegregated) covering MTM of the CDS, i.e., $L_D = \Delta_t V_t$. $L_D > 0$ when CCP posts to the bank or $L_D < 0$ when the bank posts to the CCP. The collateral is in cash earning interest at $r_L$. We assume $r_L = r$.

---

[5] Standard CDS contracts pay quarterly following the IMM convention so exact maturity match would require bespoke CDS contract. Other practicalities include bond delivery option and recovery rate as a result of dealer poll. These are not considered in this hedging exercise.
[6] Whether the CDS is CCP cleared or not, a dealer bank also posts initial margin (IM). For simplicity, we ignore IM's impact on CDS valuation and CCP default probability.

- **Bond financing account**: There is a securities financing market for corporate bonds where party B could either borrow bonds to short or borrow money to buy bonds. An exogenous constant haircut $h$ applies, $h \geq 0$. To finance the unit bond, cash is borrowed at the rate $r_p$ and with an amount of $L_B = (1-h)B_t$.

- **Debt accounts**: The segregated economy issues two classes of short term debts, $L_h$ and $N$, each having a unit price or par and being rolled at short rates $r_1(t)$ and $r_2(t)$ respectively. $L_h$ is devoted to the repo funding break, i.e., $L_h = hB$, and $N$ is residual funding of the economy, $N_t \geq 0$.

- **Economic capital (EC) reserve account**: This segregated account's sole purpose is to absorb potential losses should the reference obligator/issuer default and the segregated economy has any cash shortfall after a post-default liquidation. A reserve balance $N_c \geq 0$ is required to set aside in a separate bank account, earning interest at $r$. This amount is provided by a capital financier who requires a dividend payment at the rate of $r_k$. $\overline{r_k} = r_k - r$ is the excess return on equity.

- **Other notations**: $\bar{r}_p = (1-h)r_p + hr_1$ is the effective bond financing rate, $l(t) = B - R + \Delta_t(V - (1-R))$ is the jump-to-default exposure of the basis trade, and $\overline{M_t} = M_t - N_t, M_t = \overline{M_t}^+, N_t = \overline{M_t}^-$, combining $M_t$ and $N_t$ into one account.

Write the economy's wealth $\pi_t$ as follows,

$$\pi_t = M_t + (1-\Gamma_t)(B + \Delta_t V - L_B - L_D - L_h - N_t) = M_t - (1-\Gamma_t)N_t \quad (14)$$

Accounts excluding the bank account $M_t$ are conditional to no termination, i.e., $\Gamma_t = 0$, so that all relevant quantities are to be understood as pre-default values. To shorten the formula, all $t$-subscripts have been dropped, unless necessary. $M_t$ and $N_t$ are non-negative adapted stochastic processes. All short rates are non-stochastic by themselves, either deterministic or functions of the intensity process $\lambda$, e.g. $r_1$, $r_2$, and $r_p$.

At $t=0$, the wealth reduces to $\pi_0 = M_0 - N_0$. Let $M_0=0$, $L_{h0}=hB_0$, and $N_0=0$ so that $\pi_0 = 0$, i.e., the initial fund of the economy starts out at zero.

For $\tau > t \geq 0$, during the normal course of business, the bank pursues a trading strategy to hedge the bond and performs all necessary funding and credit support functions stipulated by the CDS and repo margin agreements. Excess cash is deposited into the bank account; debt, if any, is serviced and rolled as needed. Interests including dividend on capital are collected and/or paid.

Specifically, over a small interval of time *dt*, on the hedge front, buying $d\Delta_t$ more units of CDS protection at the price of $V_t+dV_t$ will cost cash of $d\Delta_t(V_t+dV_t)$ amount. On the collateral side, CCP posts additional collateral amount $dL_D$ in cash while being paid of interest amount $r_L L_D dt$. When the bond price increases due to credit spread tightening, the value of the CDS contracts would decline and the economy will need to return a portion of the cash collateral received as CDS margin. Because the repo account is tied up to the bond market price through a haircut, a spread tightening causing bond price to rise will increase its funding by the amount of *(1-h)dB*, which can be used to pay down debt $L_h$.

The debt accounts pay interest amount $r_1 L_h dt$ and $r_2 N dt$, roll into new issuances of $L_h+dL_h$ and $N+dN$. A negative *dN* signals a repayment of debt. The bank account accrues interest amount *rMdt*. The EC reserve account dividends out an amount of $r_k N_c dt$ while receiving interest of $r N_c dt$, a net payment of $(r_k-r)N_c dt$.

The wealth equation is written with all default effects implicitly built into the bank account $M_t$. If a default happens before *T*, i.e., $\tau < T$, trades will have settled without delay and the resulting cash flow will be swiped into the bank account which is the only account active at that point and after. $M_t$ may exhibit a jump at $\tau$ as a result of default settlement. Put everything together, the economy's pre-default financing equation follows for $t < min(T, \tau)$,

$$dM = rMdt + r_c dt + (1-h)(dB - r_p B dt) - \Delta S dt + \Delta(dV - r_L V dt) + dL_h - r_1 L_h dt + dN - r_2 N dt - (r_k - r)N_c dt \qquad (15)$$

In the last step, $L_D=\Delta_t V$ and $L_B=(1-h)B_t$ have been applied.

Upon the issuer's default at $\tau$, the bond recovers *R* fraction of its notional. CDS hedges settle at $\Delta_t (1-R)$ with the CCP (assumed non-defaultable) makes a payment of $\Delta_t ((1-R)-V(\tau))$. The repo financing account also unwinds where the defaulted obligation is returned to the bank and the repo principal and funding interest are paid to the repo buyer or lender. Adding together, the net cashflow coming to the bank account at default is $R+ \Delta_t (1-R-V(\tau))-(1-h)B(\tau)-N(\tau)$, including paying off the debts $L_h(\tau)$ and $N(\tau)$.

The risky bond fair value B in general will depend on $(M_t, N_t)$ or $\overline{M_t}$ in addition to $\lambda$. Recognizing that the CDS fair value $V$ is given by equation (9), setting the hedge ratio $\Delta_t = -\frac{\partial B}{\partial \lambda} / \frac{\partial V}{\partial \lambda}$, applying 2-D Ito's lemma involving both $\overline{M_t}$ and $\lambda$, we obtain the following PDE (see Appendix A for derivation),

$$(\frac{\partial}{\partial t}+A)B + \frac{\partial B}{\partial \overline{M}}(r\overline{M}-\lambda(R-B-(\frac{1-R-V}{\partial V/\partial \lambda})\frac{\partial B}{\partial \lambda})) \\ -\overline{r}_p B + r_c + \lambda(R-B) - (r_k-r)N_c - (r_2-r)\overline{M}^- = 0 \tag{16}$$

PDE (16) is distinctive in that it is a 2-d PDE with an additional first order derivative term (second term) on the funding account balance $\overline{M_t}$. The PDE is non-linear as it involves products of the forms: $B\frac{\partial B}{\partial \overline{M}}$ and $\frac{\partial B}{\partial \lambda}\frac{\partial B}{\partial \overline{M}}$. Its highest order derivative term however remains simple. $\overline{M_t}$ is given by

$$d\overline{M} = (r\overline{M}+\lambda l(t))dt, \text{ or}$$

$$\overline{M}_t = \beta_t^{-1}\overline{M}_0 + \beta_t^{-1}\int_0^t \lambda(u)l(u)\beta_u du \tag{17}$$

with $\overline{M}_0 = 0$. Its differential form shows that it accumulates on the cash rate and compensates the instantaneous expected loss of the trade. The integral form is path dependent, resembling the accumulated average price in an Asian option.

**Special case**: If the intensity is deterministic, we can set $\Delta_t = \frac{B-R}{1-R-V}$ to zero out the jump to default (JtD) loss $l(t)$. It is indeed hedgeable, so long as the intensity is not dynamic. Equations (16 & 17) reduce to

$$d\overline{M} = r\overline{M}dt, \\ \frac{\partial B}{\partial t} + r\overline{M}\frac{\partial B}{\partial \overline{M}} - \overline{r}_p B + r_c + \lambda(R-B) - (r_k-r)N_c - (r_2-r)\overline{M}^- = 0 \tag{18}$$

CDS PDE (9) becomes

$$\frac{dV}{dt} - (r+\lambda)V - S + \lambda(1-R) = 0 \tag{19}$$

Adding (18) and (19), noting that $\overline{M}_0 = 0$, so $\overline{M}_t = 0$, leads to

$$\frac{d(B+V)}{dt} - \bar{r}_p B - rV + r_c - S - \lambda(B+V-1) - \bar{r}_k N_c = 0 \tag{20}$$

where $\bar{r}_k = r_k - r$. $B+V$ in fact is the typical negative basis trade with the loss function $v=B+V-1$.

$v$ is not guaranteed to be zero in general. To enforce zero loss, $v=0$, bond price would have to be $B=1-V$. In this case, $\bar{M}_t = 0$, $N_c$ is zero because of no need for capital. Now if $r_c=r$ (zero spread floating rate bond) and $S=0$, equation (20) shows $\frac{dv}{dt} - (r+\lambda)v = 0$, which trivially proves that $B=1-V$ for all $t$, i.e., a zero-spread bond and zero premium CDS will always price to par in an ideal riskfree financing condition.

**A fair basis formula**: Obviously, if $\bar{r}_p = r$, then $B+V$ can be solved as one variable. In fact, it can be written in terms of $v$,

$$\frac{dv}{dt} - (r+\lambda)v + r_c - S - r - \bar{r}_k N_c = 0 \tag{21}$$

This shows that, if the bond can be financed at the riskfree rate, the negative basis trade earns a carry of $r_c$-$S$-$r$ and pays out a dividend on capital. The solution to (21), assuming constant $r_c$-$S$-$r$-$\bar{r}_k N_c$, is

$$v = apv * (r_c - S - r - \bar{r}_k N_c) \tag{22}$$

This is exactly the empirical formula some traders use to estimate their basis trade profit and loss or to evaluate whether a basis trade is economically appealing, except with the bond repo cost $\bar{r}_p$ replacing $r$, i.e., $v = apv * (r_c - S - \bar{r}_p - \bar{r}_k N_c)$.

Now setting $v$ to zero leads to a fair bond-CDS basis formula:
$$(r_c - S - z)_{fair} = (\bar{r}_p - z) + \bar{r}_k N_c. \tag{23}$$
where $z$ is a short rate for Libor. So the break-even or fair bond-CDS basis consists of the effective funding cost and cost of economic capital. The fair basis is always negative, unless the overall bond financing rate is lower than Libor[7].

---

[7] In the industry, the basis is CDS spread minus Z-spread, which is measured on top of the Libor swap curve. Both term repo funding rate and unsecured rate are quoted on top of Libor. Positive basis could also be made possible when relative liquidity favors CDS.

## 4. Economic Capital

We have one more quantity undetermined in (16), namely $N_c$, the capital account balance. Following Lou (2016), we can define a discounted hedging error (wealth) VaR measure as the economic capital. With the bond priced as such, the hedging error is left with jump risk only,

$$d\pi - r\pi dt = -l(t)(d\Gamma_t - (1-\Gamma_t)\lambda dt) \tag{24}$$

Deflating (24) and integrating to $T$ yield

$$\beta_T \pi_T - \beta_t \pi_t = \int_t^{T \wedge \tau} \beta_s l(s) \lambda ds - \beta_\tau l(\tau) I(\tau < T) \tag{25}$$

This shows that the discounted wealth of the economy accumulates continuously over time till the earlier of maturity or default, at the rate of default intensity times the jump to default loss. Now we can define the loss of economic value at time $t$ for the full remaining duration of the economy and its VaR as follows,

$$\hat{\pi}_t = \pi_t - \frac{\beta_T}{\beta_t} \pi_T$$

$$VaR_t = \inf\{x \in R : \Pr(\hat{\pi}_t > x) \leq 1 - q\}. \tag{26}$$

Because $E[\hat{\pi}_t]$ is zero, $N_c(t)=VaR_t$. When a loss is realized and the wealth is not sufficient to pay the loss, money in the reserve account will be drawn to cover loss. On an expectation basis, the wealth growth is sufficient to meet the loss, so economic capital is for unexpected losses.

As is defined, the gain/loss function $l(t)$ is coupled with the bond price through PDE (16). While a numerical solution can always be tried, there is something else we need to consider: the portfolio diversification effect. The loss function is defined on the issuer's default, in total disregard of other trading positions the bank might have. In a typical credit risk management approach, default correlation is recognized. A classic treatment is Vasicek's large loan portfolio theory where the default correlation is assumed to be constant and a limit is taken such that each constituent of the portfolio can compute its contribution to the economic capital on a standalone basis. This in fact is the foundation of BASEL II and III's wholesale credit risk capital approach (Gordy 2003).

For a unit notional exposure, BASEL III provides the following formulae to compute the regulatory capital requirement,

$$K = LGD \times [N(\frac{N^{-1}(PD) + \sqrt{\rho}N^{-1}(0.999)}{\sqrt{1-\rho}}) - PD] \times \frac{1 + (M - 2.5)b}{1 - 1.5b} \qquad (27)$$

where $K$ is the capital requirement, $PD$ is one year probability of default, $LGD$ is downturn loss given default, $M$ is the effective maturity. Additionally, $\rho$ is correlation and $b$ is a maturity adjustment factor, given as follows,

$$\rho = AVC \times [0.12 \times \frac{1 - \exp(-50 \times PD)}{1 - \exp(-50)} + 0.24 \times (1 - \frac{1 - \exp(-50 \times PD)}{1 - \exp(-50)})],$$
$$b = (0.11852 - 0.05478 \times \ln(PD))^2 \qquad (28)$$

where $AVC=1.25$ is an asset value correlation multiplier for large regulated financial institutions or unregulated financial institutions, 1 for other firms. The formulae and its parameters are further explained in BCBS 2005.

If we adopt the above as the economic capital measurement, alignment of economic capital and regulatory capital is achieved. This is advantageous in light of recent discussions of capital valuation adjustment (KVA) which seems to have become the next controversial value adjustment following FVA. KVA as advocated by some is bank specific and regulatory region dependent (Sherif, 2016). Indiscretionary application of such a KVA is certainly debatable. Consider a simple example of a trading book containing a long European option on a stock. If the trader chooses to keep it unhedged, then a market risk VaR exists. But if the trader dynamically hedges the option within the trading book, the book will have zero VaR. A trader charging KVA in the former case by lowering his bid will only cut himself off from the market and no trade will be coming to his way.

This simple example shows that for a complete market like the equity market, no KVA should be made to the fair value and charged to customers. For incomplete markets such as the credit market, the risk capital due to the systemic risk factor can't be reduced and its cost should be an integral part of pricing. Ideally, a capital charge should be based on economic capital, economically truthful to the market and its products as is. In Lou (2016), the gap risk of a secured financing transaction is analyzed and economic capital is found to be higher than regulatory capital with near zero haircuts and lower in high haircut region. KVA is then introduced as a cost of economic capital and found to play an important role in repo pricing. This type of KVA is different from regulatory capital based KVA, which has been proposed by some as an alternative to a RWA hurdle rate based pricing to help banks allocate regulatory capital costs.

Suppose we adopt equations (27 and 28) as the economic capital per unit of exposure, we then need to determine the exposure amount for the basis trade, which is given by equation (25). Although precise, this is obviously not an ideal form, because it's convoluted with the fair value of the bond and CDS. Numerical solutions can proceed, but some simpler alternatives are desired in practice. First we can adopt a fixed exposure. Second, we could establish the loss function from the risk neutral pricing of the bond and CDS fair values respectively. Third, an add-on approach can be adopted based on the hedged portfolio's CS01 and duration. Because it is a capital measure, we could always start with a more conservative approach by employing a sufficiently large fixed exposure.

## 5. Numerical Examples

PDE (16) shows that a bond's fair value not only depends on the default intensity but also relates to its accumulated default risk compensation $\overline{M}$. Once the bond's fair value $B$ is computed, it can be compared with the risk neutral bond price $B^*$ to determine a total valuation adjustment (xva), $xva = B^* - B$. Since the CDS is priced under the OIS discounting, $B^*$ is consistent with CDS price[8]. $xva$, when converted into a running spread by dividing with the risky annuity, becomes the fair basis, a measurement of fair carrying cost of a basis trade. It can be used as a relative value measure in comparison to current market implied bond-CDS basis.

---

[8] For a floating rate bond, its pricing can be obtained by considering a bond plus CDS package where the CDS references the bond and has the same spread $S_b$ as the bond's spread over the floating index. Prior to default or the bond maturity whichever comes first, the package has a net interest cash flow of the index. Upon a default or maturity, it pays the par amount. So the package's cash flow resembles that of a randomly terminating, loss free floating rate par bond that always price at the par, regardless of its termination date. Therefore the bond price $npv_{bond} = 1 - npv_{cds} = 1 - (dpv-S_b*apv) = 1 - dpv + S_b *apv$.

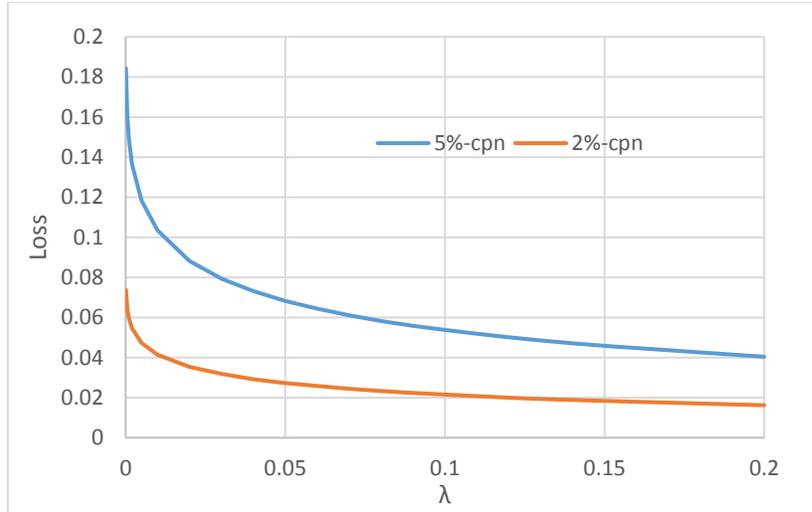

Figure 2. Sample distribution of jump-to-default loss of a basis trading strategy longing a coupon bond and a zero premium CDS protection. *r=0, R=0*.

To get some sense of the magnitude of the jump-to-loss exposure a basis trading strategy could create, we consider a 5% fixed coupon bond protected by a zero premium CDS, each priced in the usual risk neutral way. The exposure is maximized at initial time, i.e., *t=0*. Figure 2 also plots 'JtD Loss' for a 2% coupon bond. Obviously the exposure reduces significantly. The exposure is greatest at small $\lambda$, as lower $\lambda$ leads to higher bond price thus exposure at default.

A finite difference scheme is developed to solve equation (16) for the risky bond price $B$ and equation (12) for $B^*$. Implied basis between the bond and CDS can then be deduced. Table 1 shows a simple test case where a 10 year zero coupon bond and zero premium CDS are priced with the Crank-Nicholson finite difference (FD) scheme solving PDE (9&12). This test satisfies the par test, i.e., bond + CDS npv is par. Also delta and gamma are seen to be offsetting. The last row also gives the bond's price computed from 80000 paths of Monte Carlo simulation, as an accuracy check for the FD scheme.

Table 1. FD test case of bond and CDS parity vs Monte Carlo.

|      | Prc/NPV  | Delta    | Gamma     |
|------|----------|----------|-----------|
| Bond | 40.25913 | -1.09976 | 16.36011  |
| CDS  | 59.74087 | 1.09976  | -16.36011 |
| MC   | 40.23167 | -        | -         |

Consider a fixed rate coupon bond at a credit spread of 1.2%, 10 year maturity, investment grade (IG) rated. Figure 3 shows total valuation adjustment (xva) converted into basis points (bp), i.e., the fair basis, under three economic capital requirements.

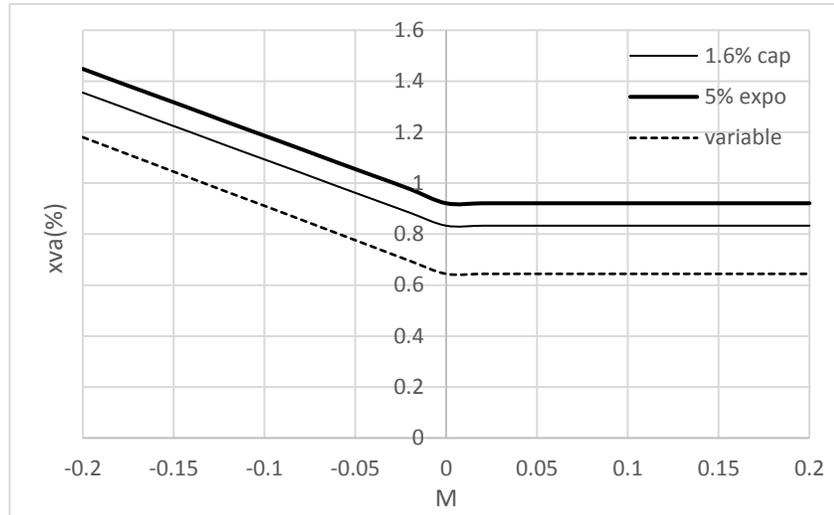

Figure 3. IG bond xva (fair basis) vs cummulative funding amount $\overline{M_t}$, with above variables in Table below, second row (IG).

| variables | T | h | $r_p$-r | $r_c$-r | $r_b$-r | r |
|---|---|---|---|---|---|---|
| IG | 10 | 10% | 0.15% | 1.20% | 1.40% | 2% |
| HY | 5 | 20% | 0.20% | 5% | 1% | 2% |

The first is a fixed 1.6% of the notional; the second is 5% fixed exposure, and the last is variable exposure by $l(t)$. In the second and last cases, the capital requirement ratio is computed per equation (27) where the PD is calculated locally as one year default probability. When the funding amount $\overline{M_t}$ turns positive (meaning deposit at the bank), xva drops as the part of funding cost corresponding to the negative $\overline{M_t}$ (for borrowing at much higher rate) disappears.

A common choice for the debt account rate $r_1$ is the dealer bank's senior unsecured rate $r_b$. For $r_2$, we mix in between $r_k$ and $r_b$, $r_2 = \varepsilon r_k + (1-\varepsilon) r_b$, where ε is the inverse of the leverage ratio, e.g., 5% given 20 time leverage. Figure 4 shows a 5 year high yield (HY) bond's total xva under the variable exposure. As $r_b$ increases, the funding component of xva increases while the capital component (KVA) remains relatively flat.

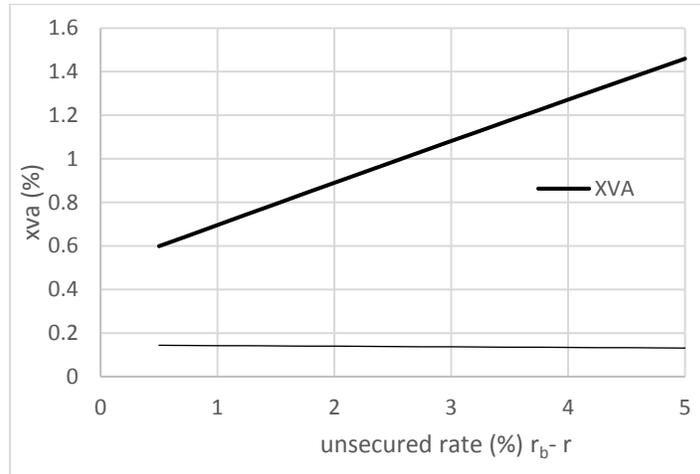

Figure 4. HY 5y 5% spread bond's xva (fair basis) vs bank's unsecured rate $r_b$. Variables in the last row of Table in Figure 3.

The levels of xva shown in Figure 3 & 4 are in the range of 60 to 120 bps, reasonably mild in the context of the historical basis graph (Figure 1.) Figure 4 in particular reflects the funding cost well through the approximate basis formula. The spike during the crisis relates to funding and market volatility. Table 2 lists the average basis for the two halves of 2007 and the whole 2008, together with overnight repo spreads to the Fed funds rate with 'BBB' and 'A' rated US corporate bond collateral. Clearly, widening of the basis accompanied increased repo spread and market volatility index VIX, as the crisis progressed, and large negative basis trading losses were reported.

Table 2. Average IG and HY basis for 1st and 2nd half of 2017 and year 2008, with bond funding costs and VIX (from Gorton and Metric, 2012).

|  | repo spread (bp) | VIX(%) | IG Basis | HY Basis |
|---|---|---|---|---|
| 1st Half 2007 | 2.01 | 13.05 | -1.32 | 15.3 |
| 2nd Half 2007 | 61.85 | 21.88 | -19.65 | -4.2 |
| 2008 | 136.19 | 33.68 | -103.26 | -210.2 |

Using three month Libor-OIS spread (LOIS) as a control variable for funding cost and VIX for capital, regression of the daily time series from Jan 2005 to March 2018 (Table 3) shows that both LOIS and VIX are significant. They explain about 65% of variation, for both the IG and HY historical basis. Noting that VIX is an equity market volatility, not exactly a measure of credit risk capital and that market liquidity variable has not been included, this leaves room for future research.

Table 3. Basis regression results (coefficient column followed by standard error column) vs Libor-OIS spread and VIX. IG $R^2$ 0.648, HY $R^2$ 0.662.

|  | IG | IG-error | HY | HY-error |
| --- | --- | --- | --- | --- |
| Intercept | 0.34564 | 0.012644 | 1.0125 | 0.027542 |
| LOIS | -46.341 | 2.5188 | -45.642 | 4.7025 |
| VIX | -2.6624 | 0.082811 | -7.9847 | 0.18038 |

## 6. Conclusions

The basis is interesting or rather naughty in that it is a theoretical arbitrage opportunity, yet arbitragers often incur large losses. We propose a risky bond valuation model that understands its behavior from a funding and capital perspective. Funding cost comes from bond financing which is not riskfree. As the credit spread is dynamic, the basis trade can only hedge its diffusion risk and has to leave its residual jump to default risk unhedged. Because corporate default is known to be correlated, this residual default risk is not diversifiable. An economic capital reserve has to be taken and that incurs capital cost.

These cost factors, not considered in the traditional reduced form risk neutral pricing model, lower a defaultable bond's valuation, thus contributing to the negative basis. By setting up a risky bond dynamically hedged by CCP cleared CDS, we derive a two dimensional, quasi-linear PDE for the basis trade. Vasicek's large loan portfolio theory is used to estimate economic capital, in line with BASEL's wholesale credit risk capital requirement. The fair value solved from the PDE via finite difference method is converted into a fair basis to facilitate assessment of basis trade economics.

Our approach to cost of capital differs from KVA, a firm specific and regime sensitive regulatory capital value adjustment advocated by some researchers. We in fact imply a different form of KVA by associating it with unhedged systemic default risk and its resultant economic capital.

Numerical computation shows that reasonable repo financing cost and capital charge can adequately predict the negative basis. Limited basis data regression analysis confirms the role of these factors, although further empirical research will be desired to fully explain the basis.

**Appendix A**. Risky Bond Valuation PDE Derivation

Differentiate equation (14) and plug in $dM_t$ from (15),

$$d\pi - r\pi dt = (1 - \Gamma_t)(r_c dt + dB - ((1-h)r_p + hr_1)Bdt + \Delta(dV - r_L Vdt - Sdt) - (r_2 - r)Ndt - (r_k - r)N_c dt) - l(t)d\Gamma_t \quad (A.1)$$

with $l(t) = B - R + \Delta(V - (1-R))$. Complementing $d\Gamma_t$ with its compensator, (A.1) becomes,

$$d\pi - r\pi dt = (1 - \Gamma_t)[dB - \bar{r}_p Bdt + r_c dt + \lambda dt(R - B) - (r_2 - r)Ndt - (r_k - r)N_c dt + \Delta(dV - r_L Vdt - Sdt + \lambda dt(1 - R - V))] - l(t)(d\Gamma_t - (1 - \Gamma_t)\lambda dt) \quad (A.2)$$

Pre-default, rewrite (15) as follows,

$$d(M - N) - r(M - N)dt = r_c dt + dB - \bar{r}_p Bdt + \Delta(dV - r_L Vdt - Sdt) - (r_2 - r)Ndt - (r_k - r)N_c dt \quad (A.3)$$

Combining $M$ and $N$ into one account $\overline{M}_t$, with $\overline{M}_0 = 0$, equation (A.3) becomes

$$d\overline{M} - r\overline{M}dt = r_c dt + dB - \bar{r}_p Bdt - (r_2 - r)\overline{M}^- dt - (r_k - r)N_c dt + \Delta(dV - r_L Vdt - Sdt) \quad (A.4)$$

In general, $B_t$ could be a function of $\lambda_t$ and $\overline{M}_t$, so that a 2-dimensional Ito's lemma applies,

$$dB = \frac{\partial B}{\partial t}dt + \frac{\partial B}{\partial \lambda}d\lambda + \frac{\partial B}{\partial \overline{M}}d\overline{M} + \frac{1}{2}\frac{\partial^2 B}{\partial \lambda^2}<d\lambda, d\lambda> + \frac{1}{2}\frac{\partial^2 B}{\partial \overline{M}^2}<d\overline{M}, d\overline{M}> + \frac{\partial^2 B}{\partial \lambda \partial \overline{M}}<d\lambda, d\overline{M}>$$

We are looking for a solution where $d\overline{M}_t$ has $dt$ term but no diffusion term, so that the last two variations are zero. In such a case, the financing equation (21) becomes,

$$d\overline{M} - r\overline{M}dt = dt[r_c + (\frac{\partial}{\partial t} + A)B + \frac{\partial B}{\partial \overline{M}}d\overline{M}/dt - \bar{r}_p B - (r_2 - r)\overline{M}^- - (r_k - r)N_c] + \Delta dt[(\frac{\partial}{\partial t} + A)V - r_L V - S] \quad (A.5)$$

where $\Delta = -\frac{\partial B}{\partial \lambda}/\frac{\partial V}{\partial \lambda}$, and $A$ is short for $a\frac{\partial}{\partial \lambda} + \frac{1}{2}b^2\frac{\partial^2}{\partial \lambda^2}$.

Now back to the wealth equation (A.2), apply 2-D Ito's lemma to $dB_t$ and $V$'s PDE (9), set the hedge ratio $\Delta_t = -\frac{\partial B}{\partial \lambda} / \frac{\partial V}{\partial \lambda}$, and rewrite equation (A.2) as follows

$$d\pi - r\pi dt = (1-\Gamma_t)dt[(\frac{\partial}{\partial t} + A)B + \frac{\partial B}{\partial \overline{M}}\frac{d\overline{M}}{dt} - \bar{r}_p B + r_c + \lambda(R-B)$$
$$-(r_2 - r)\overline{M}^- - (r_k - r)N_c] - l(t)(d\Gamma_t - (1-\Gamma_t)\lambda dt) \tag{A.6}$$

To make the discounted wealth $\beta_t \pi_t$ a martingale, setting $dt$ term to zero leads to

$$(\frac{\partial}{\partial t} + A)B + \frac{\partial B}{\partial \overline{M}}\frac{d\overline{M}}{dt} - \bar{r}_p B + r_c + \lambda(R-B) - (r_2 - r)\overline{M}^- - (r_k - r)N_c = 0 \tag{A.7}$$

Equation (A.6) reduces to (24). Given this, once again, the financing equation (A.5) reduces to,

$$d\overline{M} = (r\overline{M} + \lambda l(t))dt \tag{A.8}$$

Indeed, $d\overline{M}$ has only $dt$ term, no diffusion term, and $d\overline{M}/dt = r\overline{M} + \lambda l(t)$. (A.7) now becomes equation (16).